\PassOptionsToPackage{unicode}{hyperref}
\PassOptionsToPackage{hyphens}{url}
\PassOptionsToPackage{dvipsnames,svgnames,x11names}{xcolor}
\documentclass[
]{article}
\usepackage{amsmath,amssymb}
\usepackage{lmodern}
\usepackage{iftex}
\usepackage{parcolumns}
\ifPDFTeX
  \usepackage[T1]{fontenc}
  \usepackage[utf8]{inputenc}
  \usepackage{textcomp} 
\else 
  \usepackage{unicode-math}
  \defaultfontfeatures{Scale=MatchLowercase}
  \defaultfontfeatures[\rmfamily]{Ligatures=TeX,Scale=1}
\fi
\IfFileExists{upquote.sty}{\usepackage{upquote}}{}
\IfFileExists{microtype.sty}{
  \usepackage[]{microtype}
  \UseMicrotypeSet[protrusion]{basicmath} 
}{}
\makeatletter
\@ifundefined{KOMAClassName}{
  \IfFileExists{parskip.sty}{%
    \usepackage{parskip}
  }{
    \setlength{\parindent}{0pt}
    \setlength{\parskip}{6pt plus 2pt minus 1pt}}
}{
  \KOMAoptions{parskip=half}}
\makeatother
\usepackage{xcolor}
\usepackage{graphicx}
\makeatletter
\def\maxwidth{\ifdim\Gin@nat@width>\linewidth\linewidth\else\Gin@nat@width\fi}
\def\maxheight{\ifdim\Gin@nat@height>\textheight\textheight\else\Gin@nat@height\fi}
\makeatother
\setkeys{Gin}{width=\maxwidth,height=\maxheight,keepaspectratio}
\makeatletter
\def\fps@figure{htbp}
\makeatother
\NewDocumentCommand\citeproctext{}{}
\NewDocumentCommand\citeproc{mm}{%
  \begingroup\def\citeproctext{#2}\cite{#1}\endgroup}
\makeatletter
 \let\@cite@ofmt\@firstofone
 \def\@biblabel#1{}
 \def\@cite#1#2{{#1\if@tempswa , #2\fi}}
\makeatother
\newlength{\cslhangindent}
\newlength{\csllabelwidth}
\newenvironment{CSLReferences}[2] 
 {\begin{list}{}{%
  \setlength{\itemindent}{0pt}
  \setlength{\leftmargin}{0pt}
  \setlength{\parsep}{0pt}
  \ifodd #1
   \setlength{\leftmargin}{\cslhangindent}
   \setlength{\itemindent}{-1\cslhangindent}
  \fi
  \setlength{\itemsep}{#2\baselineskip}}}
 {\end{list}}
\usepackage{calc}

\ifLuaTeX
\usepackage[bidi=basic]{babel}
\else
\usepackage[bidi=default]{babel}
\fi
\babelprovide[main,import]{american}

\def\languageshorthands#1{}
\ifLuaTeX
  \usepackage{selnolig}  
\fi
\IfFileExists{bookmark.sty}{\usepackage{bookmark}}{\usepackage{hyperref}}
\IfFileExists{xurl.sty}{\usepackage{xurl}}{} 
\hypersetup{
  pdftitle={yieldplotlib: A unified library for exoplanet yield code
visualizations},
  pdfauthor={Corey Spohn, Sarah Steiger, Alex R. Howe},
  pdflang={en-US},
  colorlinks=true,
  linkcolor={Maroon},
  filecolor={Maroon},
  citecolor={Blue},
  urlcolor={Blue},
  pdfcreator={LaTeX via pandoc}}

\title{yieldplotlib: A unified library for exoplanet yield code
visualizations}

\usepackage[affil-it]{authblk}
\usepackage{orcidlink}
\author[1%
  *%
  ]{Corey Spohn%
    \,\orcidlink{0000-0002-3138-0240}\,%
    }
\author[2%
  *%
  ]{Sarah Steiger%
    \,\orcidlink{0000-0002-4787-3285}\,%
    }
\author[1,3,4%
  ]{Alex R. Howe%
    \,\orcidlink{0000-0002-4884-7150}\,%
    }

\affil[1]{Goddard Space Flight Center, United States%
  }
\affil[2]{Space Telescope Science Institute, United States%
  }
\affil[3]{The Catholic University of America, United States%
  }
\affil[4]{Center for Research and Exploration in Space Science and
Technology, NASA/GSFC, United States%
  }
\date{1 April 2025}

\begin{document}
\maketitle

\def\thefootnote{*}
\footnotetext{These authors contributed equally.}
\section{Summary}\label{summary}

NASA's next flagship observatory, the Habitable Worlds Observatory
(HWO), aims to detect and charaterize \textasciitilde25 habitable zone
planets. The total number of habitable zone planets detected is referred
to as the exo-Earth ``yield'' and accurate yield estimates will be
critically important to the mission's success. Tools like the Altruistic
Yield Optimizer (AYO) and EXOSIMS provide these yield estimates but
differ in language, methods, and outputs. \texttt{yieldplotlib} provides
a unified library that can visualize the inputs and outputs of these
yield codes in a complete, descriptive, and accessible way. To learn more about \texttt{yieldplotlib}, please visit the \href{https://github.com/HWO-Yield-Visualizations/yieldplotlib}{GitHub} repository or see the documentation hosted on \href{https://yieldplotlib.readthedocs.io/en/latest/user/intro.html}{Read the Docs}. 

\section{Statement of need}\label{statement-of-need}

To evaluate different designs for HWO
(\citeproc{ref-HWOFeinberg2024}{Feinberg et al., 2024}), yield codes
such as AYO (\citeproc{ref-AYO2014}{Stark et al., 2014}) and EXOSIMS
(\citeproc{ref-EXOSIMS2016}{Delacroix et al., 2016}) calculate the
expected exo-Earth yield for each architecture. While these yield codes
have the same goal, their implementations are so different that
validation has become a major obstacle for the community.

Previous cross-calibration efforts have already provided valuable
insights. Stark et al. (\citeproc{ref-ETCCrossCal2025}{2025}) compared
the internal exposure time calculations of AYO and EXOSIMS and revealed
previously unknown discrepancies. \texttt{yieldplotlib} is a more
ambitious continuation of that work. It is a Python library capable of
easily accessing hundreds of important quantities from both AYO and
EXOSIMS.

To visualize the inputs and outputs of AYO and EXOSIMS,
\texttt{yieldplotlib} uses a custom loading and parsing structure to
easily access equivalent data across both codes. This allows
\texttt{yieldplotlib} to communicate the results of yield codes to the
broader community and produce publication-quality plots without manually
processing the complex underlying data. Currently \texttt{yieldplotlib}
contains modules for analyzing AYO and EXOSIMS, but can be easily
extended to support other yield codes.

\section{Methods and Functionality}\label{methods-and-functionality}

\subsection{Parsing and Getting
Values}\label{parsing-and-getting-values}

\begin{figure}
\centering
\includegraphics[keepaspectratio]{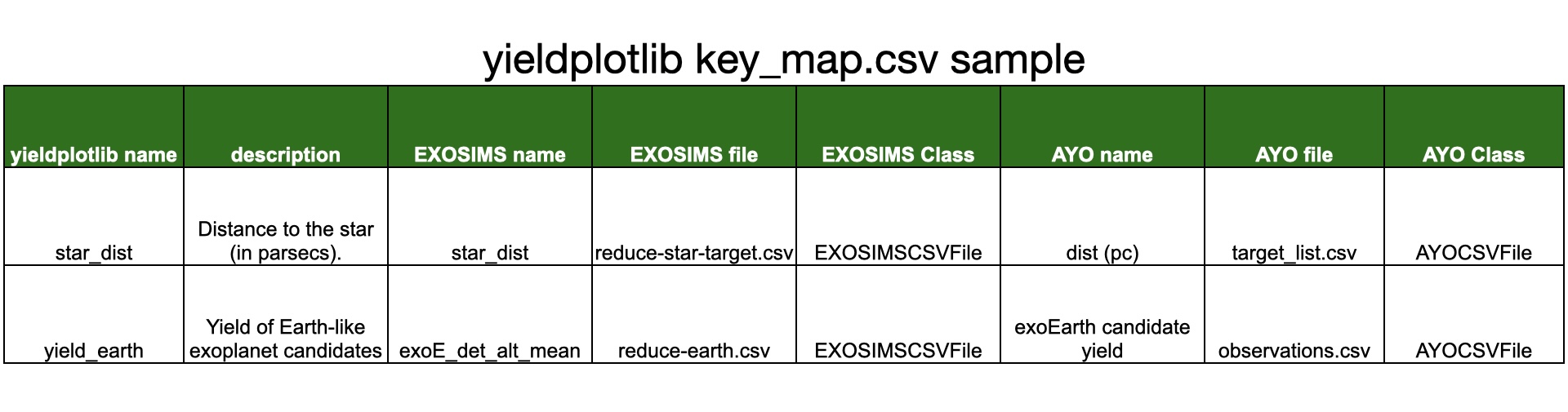}
\caption{Example portion of the \texttt{yieldplotlib} key map containing
the mappings between AYO, EXOSIMS, and \texttt{yieldplotlib}
parameters.\label{fig:key_map_csv}}
\end{figure}

\texttt{yieldplotlib} provides a loading system with a unified interface
for accessing data from the yield codes. The system internally manages
the complex and inconsistent file structures of the AYO and EXOSIMS
inputs and outputs so that users can access data with simple and
consistent queries. For collaboration purposes, the valid queries are
managed in a Google Sheet (see \autoref{fig:key_map_csv}) and
automatically processed into a universal key map.

\subsection{Plotting}\label{plotting}

\subsubsection{Generic and Comparison
Plots}\label{generic-and-comparison-plots}

\texttt{yieldplotlib} extends the widely used Python plotting library
\texttt{matplotlib}, leveraging its extensive customization options and
the familiarity many users will already have with it. The
\texttt{yieldplotlib} generic plots (\texttt{ypl\_plot},
\texttt{ypl\_scatter}, and \texttt{ypl\_hist}) are used for single yield
run visualizations.

To compare multiple yield runs, \texttt{yieldplotlib} also provides a
set of flexible comparison plots that create multi-panel figures to
quickly identify discrepancies.

In order to generate summary plots quickly, \texttt{yieldplotlib}
provides a command line interface and plotting pipeline to create a
suite of commonly used yield plots.

\subsubsection{Plotting Scripts}\label{plotting-scripts}

\texttt{yieldplotlib} contains scripts for generating common plots used
in yield code visualizations. These allow users to instantly compare AYO
and EXOSIMS results as motivated by the rapid progress of HWO's ongoing
architecture trade studies. These scripts also function as examples for
users who want to adapt the generic \texttt{yieldplotlib} methods to
generate bespoke visualizations.

\begin{figure}
\centering
\includegraphics[keepaspectratio]{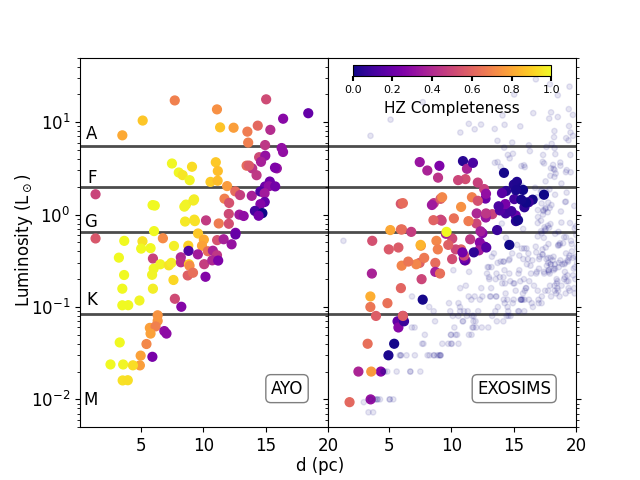}
\caption{Plot of the Habitable Zone (HZ) completeness as a function of
host star luminosity (in units of Solar luminosity) and distance (in
parsecs). Here the AYO results are on the left and the EXOSIMS results
are on the right.\label{fig:hz_completeness}}
\end{figure}

\begin{figure}
\centering
\includegraphics[keepaspectratio]{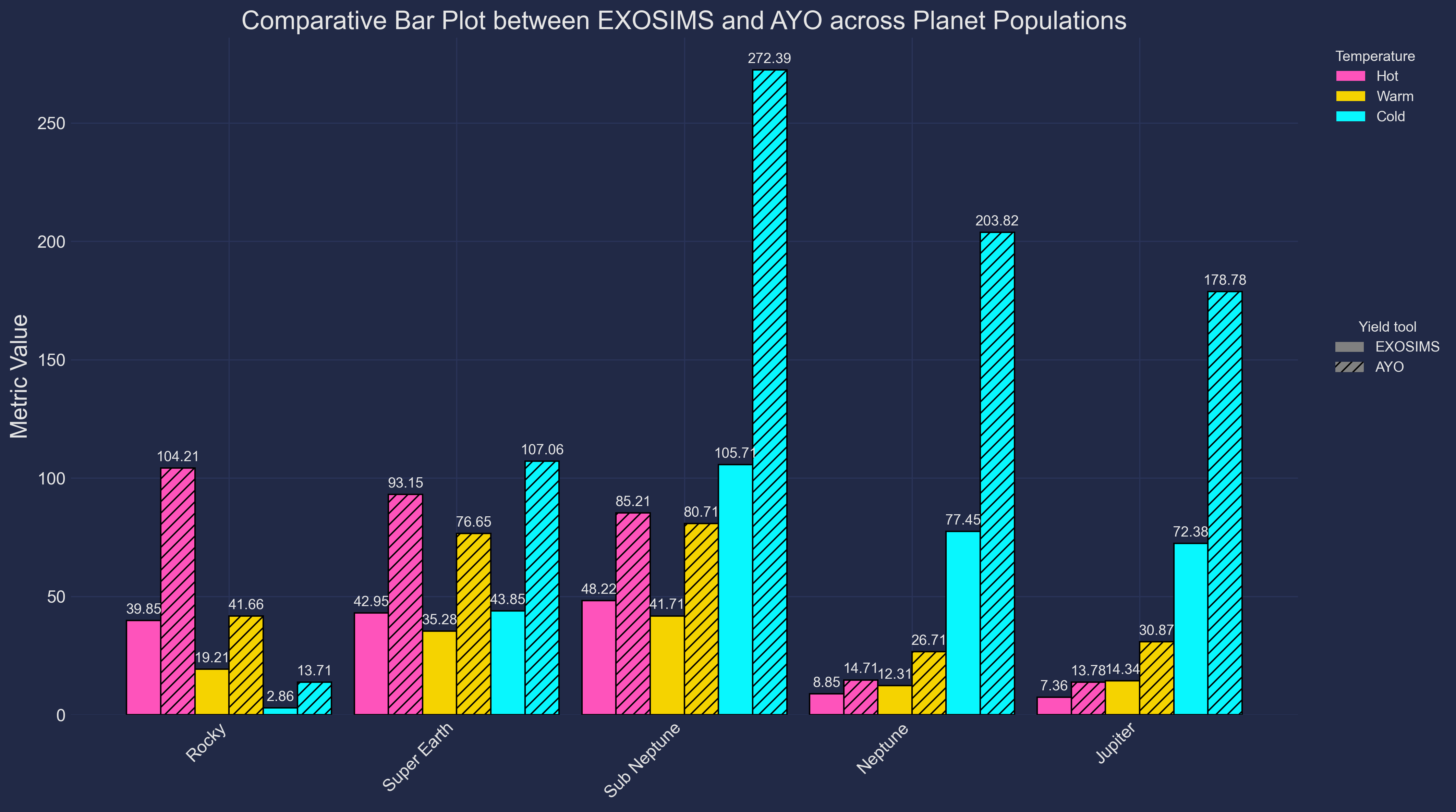}
\caption{Bar chart comparing AYO and EXOSIMS planet yields for different
classes of planets. This plot demonstrates \texttt{yieldplotlib}'s usage
of the \texttt{mplcyberpunk} color scheme as an alternative dark mode
style. Note that this is a demonstration plot only, the AYO and EXOSIMS
inputs shown are not directly comparable. \label{fig:planet_hists}}
\end{figure}

\autoref{fig:hz_completeness} compares the mission's ``habitable zone
completeness'', the probability that the simulated mission's
observations would detect a planet in the habitable zone if one exists,
as calculated by the two yield codes side-by-side.
\autoref{fig:planet_hists} shows histograms of the total number of
detected planets found as a function of planet type for the two codes
though, in this example, the inputs to each code also differ resulting
in the seen discrepancies.

\begin{figure}
\centering
\includegraphics[keepaspectratio]{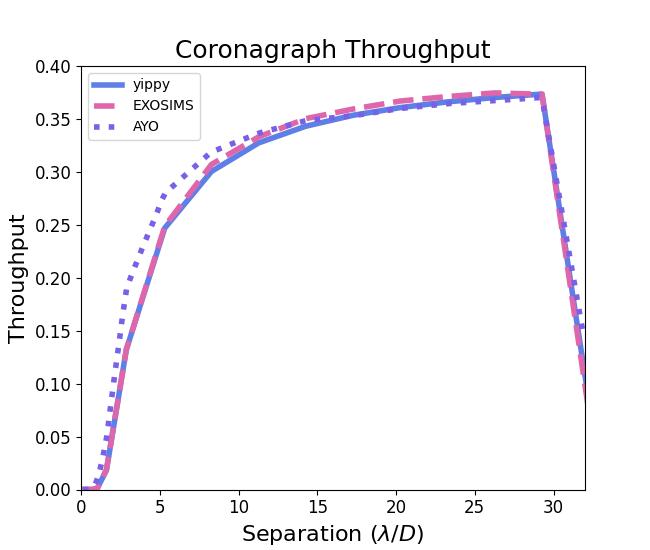}
\caption{Core throughput vs.~separation (in \(\lambda\)/D) for the same
coronagraph when processed by AYO, EXOSIMS, and calculated by a tool
named \texttt{yippy}. The differences are due to the different
interpolation methods used and the definition of aperture over which the
throughput is calculated. This highlights the types of insights that
tools like \texttt{yieldplotlib} can help to uncover.
\label{fig:core_throughput}}
\end{figure}

Yield code inputs have a profound impact on calculated yield and
plotting them is important to ensure consistency. Yield input packages
(YIPs), a set of files that describe coronagraph performance, can also
be loaded and accessed in \texttt{yieldplotlib} to compare how different
codes process the same input coronagraph. \autoref{fig:core_throughput}
shows a comparison of the calculated coronagraph throughput for both the
two codes studied, and as accessed via an additional YIP analysis tool
called \texttt{yippy}. Smaller throughputs result in less planet light
on the detector which can result in lower yields.

\section{Future Work}\label{future-work}

A new cross-calibration study of yield codes is being organized and will
utilize \texttt{yieldplotlib}. This study will be vital to the ongoing
HWO architecture trade studies and ensure more reliable and robust yield
estimates.

\section{Acknowledgements}\label{acknowledgements}

Corey Spohn's research was supported by an appointment to the NASA
Postdoctoral Program at the NASA Goddard Space Flight Center,
administered by Oak Ridge Associated Universities under contract with
NASA. Sarah Steiger acknowledges support from an STScI Postdoctoral
Fellowship.

The authors would also like to acknowledge Christopher Stark, Dmitry
Savransky, Rhonda Morgan, and Armen Tokadjian for providing consultation
on the AYO and EXOSIMS repositories. They would also like to thank
Justin Hom, and the rest of the Exoplanet Science Yields Working Group
(ESYWG) for their valuable feedback and discussions as well as Susan
Redmond for the development of the sample Yield Input Package.

\end{document}